\documentstyle{camera}

\newcommand{\AmS}{{\protect\the\textfont2
  A\kern-.1667em\lower.5ex\hbox{M}\kern-.125emS}}

\newcommand{\beq}{\begin{equation}}
\newcommand{\eeq}{\end{equation}}
\newcommand{\bea}{\begin{eqnarray}}
\newcommand{\eea}{\end{eqnarray}}

\def\dm2{\Delta m^2}
\def\sq2{sin^2(2\Theta)}

\def \SAIT #1 #2 {{\em Mem.\ Soc.\ Astron.\ It.\/} {\bf #1}, #2}
\def \MESS #1 #2 {{\em The Messenger\/} {\bf #1}, #2}
\def \ASTRNACH #1 #2 {{\em Astron. Nach.\/} {\bf #1}, #2}
\def \AA #1 #2 {{\em Acta Astron.\/} {\bf #1}, #2}
\def \AAP #1 #2 {{\em Astron. Astrophys.\/} {\bf #1}, #2}
\def \AAL #1 #2 {{\em Astron. Astrophys. Lett.\/} {\bf #1}, L#2}
\def \AAR #1 #2 {{\em Astron. Astrophys. Rev.\/} {\bf #1}, #2}
\def \AAS #1 #2 {{\em Astron. Astrophys. Suppl. Ser.\/} {\bf #1}, #2}
\def \AJ #1 #2 {{\em Astron. J.\/} {\bf #1}, #2}
\def \ANNREV #1 #2 {{\em Ann. Rev. Astron. Astrophys.\/} {\bf #1}, #2}
\def \APJ #1 #2 {{\em Astrophys. J.\/} {\bf #1}, #2}
\def \APJL #1 #2 {{\em Astrophys. J. Lett.\/} {\bf #1}, L#2}
\def \APJS #1 #2 {{\em Astrophys. J. Suppl.\/} {\bf #1}, #2}
\def \APSS #1 #2 {{\em Astrophys. Space Sci.\/} {\bf #1}, #2}
\def \ASR #1 #2 {{\em Adv. Space Res.\/} {\bf #1}, #2}
\def \BAIC #1 #2 {{\em Bull. Astron. Inst. Czechosl.\/} {\bf #1}, #2}
\def \JSQRT #1 #2 {{\em J. Quant. Spectrosc. Radiat. Transfer\/} {\bf #1}, #2}
\def \MN #1 #2 {{\em Mon. Not. R. Astr. Soc.\/} {\bf #1}, #2}
\def \MEM #1 #2 {{\em Mem. R. Astr. Soc.\/} {\bf #1}, #2}
\def \PLR #1 #2 {{\em Phys. Lett. Rev.\/} {\bf #1}, #2}
\def \PASJ #1 #2 {{\em Publ. Astron. Soc. Japan\/} {\bf #1}, #2}
\def \PASP #1 #2 {{\em Publ. Astr. Soc. Pacific\/} {\bf #1}, #2}
\def \NAT #1 #2 {{\em Nature\/} {\bf #1}, #2}
\def \SSREV #1 #2 {{\em Space Sci. Rev.\/} {\bf #1}, #2}
\def \aph #1 {{\em astro-ph\/} #1}
\def \IAUC #1 {{\em IAUC} #1}
\def \CJAA #1 #2 {{\em Chinese J. Astron. Astrophys.\/} {\bf #1}, #2}
\def \MSAIT #1 #2 {{\em Mem. Soc. Astron. Ital.\/} {\bf #1}, #2}
\def \ATEL #1 {{\em Astron. Tel.} #1}
\def \AIPC #1 #2 {{\em American Inst. Phys. Conf.\/} {\bf #1}, #2}

\input epsf

\begin{document}

%
\title{ON THE APPARENT LACK OF Be X-RAY BINARIES WITH BLACK HOLES}

%
\author{JANUSZ ZI\'O{\L}KOWSKI$^1$ \And KRZYSZTOF BELCZY\'NSKI$^2$}

%
\organization{$^1$Copernicus Astronomical Center\\ ul. Bartycka 18,
00-716 Warsaw, Poland\\$^2$Astronomical Observatory, University of
Warsaw\\ Al. Ujazdowskie 4, 00-478 Warsaw, Poland}

\maketitle

\begin{abstract}

In the Galaxy there are $72$ Be X-ray binaries known to-date. Out of
those, $48$ host a neutron star, and for the reminder the nature of
a companion is not known. None, so far, is known to host a black
hole. This disparity is referred to as a missing Be -- black hole
X-ray binary problem. The stellar population synthesis calculations
following the formation of Be X-ray binaries (Belczy\'nski \&
Zi\'o{\l}kowski 2009) predict that the ratio of the binaries with
neutron stars to the ones with black holes is rather high $F_{\rm
NS/BH} \sim 30-50$. The ratio is a natural outcome of {\em (1)} the
stellar initial mass function that provides more neutron stars than
black holes and {\em (2)} common envelope evolution (i.e.~a major
mechanism involved in the formation of interacting binaries) that
naturally selects progenitors of Be X-ray binaries with neutron
stars (comparable mass binaries have more likely survival
probabilities) against ones with black holes (much more likely
common envelope mergers). A comparison of this ratio with the number
of confirmed Be -- neutron star X-ray binaries ($48$) indicates that
the expected number of Be -- black hole X-ray binaries is of the
order of only $\sim 0-2$. This is entirely consistent with the
observed Galactic sample. Therefore, there is no problem of the
missing Be+BH X-ray binaries for the
Galaxy\\

In the Magellanic Clouds there are $98$ Be X-ray binaries known
to-date. Out of those, $63$ host a neutron star. Again, none hosts a
black hole. The stellar population synthesis calculations carried
out specifically for the Magellanic Clouds (Zi\'o{\l}kowski \&
Belczy\'nski 2010) predict that the ratio of the Be X-ray binaries
with neutron stars to the ones with black holes is only $F_{\rm
NS/BH} \sim 10$. This value is rather too low, as it implies the
expected number of Be+BH X-ray binaries of the order of $\sim 6$,
while none is observed. We believe, that to remove the discrepancy,
one has to take into account a different history of the star
formation rate in the Magellanic Clouds, with the respect to the
Galaxy. New stellar population synthesis calculations are currently
being carried out.

An updated (as of November 2011) list of all 170 Be X-ray binaries
known presently in the Galaxy and in the Magellanic Clouds is
included.

\end{abstract}
\vspace{1.0cm}

\section{Introduction}

High mass X-ray binaries host a compact object (a neutron star or a
black hole) and a massive star. The major subclass of high mass
X-ray binaries consists of a Be star and a compact object and they
are referred to as Be X-ray binaries (Be XRBs). The Be stars are
massive, generally main sequence, stars of spectral types A0-O8 with
Balmer emission lines (Negueruela 1998). The Be XRBs are found with
rather wide (orbital periods in the range of $\sim 10-350$ days) and
frequently eccentric orbits and a compact object accretes from the
wind of a Be star (even massive Be stars are within their Roche
lobes for these wide orbits). At present, $170$ Be XRBs are known in
the Galaxy and in the Magellanic Clouds, and in $111$ of them, the
compact object was confirmed to be a neutron star (NS) by the
presence of the X-ray pulsations. In the remaining cases, whenever
we have information concerning the nature of the compact component
(such as an X-ray spectrum), it also indicates a NS. Although one
cannot exclude that a few of these systems contain white dwarfs or
black holes, it is fair to state that majority of them contain NSs
as compact components. We know, at present, $60$ black hole
candidate systems in the Galaxy and in the Magellanic Clouds (among
them 22 confirmed BH systems; e.g., Remillard \& McClintock 2006;
Zi\'o{\l}kowski 2008). However, not a single black hole binary
containing a Be type component has been found so far. This
disparity, $111$ Be XRBs with NSs versus not a single one with a BH,
seems indeed striking.

The X-ray emission from Be XRBs (with a few exceptions) is of a
distinctly transient nature with rather short active phases
separated by much longer quiescent intervals (a flaring behavior).
There are two types of flares, which are classified as Type I
outbursts (smaller and regularly repeating) and Type II outbursts
(larger and irregular; Negueruela \& Okazaki 2001, Negueruela et
al.~2001). Type I bursts are observed in systems with highly
eccentric orbits. They occur close to periastron passages of a NS.
They are repeating at intervals $\sim P_{\rm orb}$. Type II bursts
may occur at any orbital phase. They are correlated with the
disruption of the excretion disc around Be star (as observed in
H$\alpha$ line). They repeat on time scale of the dynamical
evolution of the excretion disc ($\sim$ few to few tens of years).
This recurrence time scale is generally much longer than the orbital
period (Negueruela et al.~2001).

Be XRBs systems are known to contain two discs: excretion disc
around Be star and accretion disc around neutron star. Both discs
are temporary: excretion disc disperses and refills on time scales
$\sim$ few to few tens of years (dynamical evolution of the disc,
formerly known as the ``activity of a Be star'' (Negueruela et
al.~2001)), while the accretion disc disperses and refills on time
scales $\sim$ weeks to months (which is related to the orbital
motion on an eccentric orbit and, on some occasions, also to the
major instabilities of the other disc). The accretion disc might be
absent over a longer period of time ($\sim$ years), if the other
disc is very weak or absent. The X-ray emission of Be XRBs binaries
is controlled by the centrifugal gate mechanism, which, in turn, is
operated both by the periastron passages (Type I bursts) and by the
dynamical evolution of the excretion disc (both types of bursts).
This mechanism explains the transient nature of the X-ray emission (
see Zi\'o{\l}kowski 2002 and references therein).

The more detailed description of the properties of Be XRBs systems
is given, e.g. in Negueruela et al. 2001,  Zi\'o{\l}kowski 2002,
Belczy\'nski \& Zi\'o{\l}kowski 2009 and references therein.

The list of all presently known Be XRBs in the Galaxy is given in
Table 1, and the list of those known in the Magellanic Clouds is
given in Table 2.

In further discussion, I will present the recent stellar population
synthesis (SPS) calculations (Belczy\'nski \& Zi\'o{\l}kowski 2009,
Zi\'o{\l}kowski \& Belczy\'nski 2010) aimed at the understanding of
the origins of the apparent disparity of the number of known Be XRBs
with neutron stars (NSs) ($111$) as compared to no known Be XRBs
with black holes (BHs) in the Galaxy and in the Magellanic Clouds.
This disparity is referred to as a missing Be -- black hole X-ray
binary problem.

\section{SPS Calculations}

\subsection{SPS Code}

We evolve a population of massive binaries using {\tt StarTrack}
stellar population synthesis code (Belczy\'nski, Kalogera \& Bulik
2002 and Belczy\'nski et al. 2008). We adopt a steep initial mass
function (IMF) for massive stars with a power-law exponent of $-2.7$
(Kroupa \& Weidner 2003). We adopt solar metallicity ($Z=0.02$) for
galactic binaries and low metallicity ($Z=0.008$) for Magellanic
Clouds binaries. Roche lobe overflow is treated in a
non-conservative way (with $50\%$ mass loss from a given binary;
e.g.~Meurs \& van den Heuvel 1989) while the CE phase is treated via
energy balance with fully efficient transfer of orbital energy into
dispersal of an envelope (e.g.~$\alpha \times \lambda =1.0$). The
results are calibrated in such a way that the Galactic star
formation rate is at the level of 3.5 M$_{\odot}$/yr and is constant
through the last 10~Gyr (e.g.~O'Shaughnessy et al. 2008). At the
present Galactic disk age ($t=10$ Gyr) we perform a time slice and
extract Be X-ray binaries using classification criteria defined in
the following section.

\subsection{Definition of a Be XRB for the purpose of SPS
calculations}

During our SPS calculations, we consider any system a Be X-ray
binary if: {\em (1)} it hosts either a NS or a BH accretor; {\em
(2)} donor is a main sequence star (burning H in its core); {\em
(3)} donor mass is higher than 3 M$_{\odot}$ (O/B star); {\em (4)}
orbital period is in the range $10 \leq P_{\rm orb} \leq 300$ day;
and {\em (5)} only a fraction $F_{\rm Be}=0.25$ of the above systems
are designated as hosting a Be star and not a regular O/B star.

The last condition is based on the observations indicating that the
fraction of Be stars among all B stars is $1/5$ to $1/3$ (e.g.,
Zi\'o{\l}kowski 2002; McSwain \& Gies 2005).

Our set of conditions means that we assume that whenever donor is a
Be star then the accretion is always efficient, independently of the
size of the binary orbit (as is, in fact, observed in Be/NS XRBs).

\subsection{SPS Models}

We carried out the calculations for three models of SPS. In model A,
it was assumed that the binary system will survive the situation
when the donor star will overflow its Roche lobe while crossing the
Hertzsprung gap. With the present state of knowledge, it seems
doubtful, as this would rather lead to a merger of both components
(Taam \& Sandquist 2000, Ivanova \& Taam 2004). However, since model
A used to be a standard in the past, we still carried out the
calculations for this case. More realistic seem to be models B and
C, which assume that overflowing by a donor its Roche lobe while
crossing the Hertzsprung gap, leads to a merger and removal of the
binary from the statistics. The difference between the models B and
C concerns the natal kicks compact objects receive at birth. Model B
assumes for NSs the kicks drawn from the radio pulsar birth velocity
distribution derived by Hobbs et al. (2005; a Maxwellian with
$\sigma=265$~km/s). However, there are some indications that natal
kicks neutron stars receive are smaller for stars in binaries as
compared to single stars (e.g.~Podsiadlowski et al.~2004).
Therefore, model C assumes a Maxwellian distribution with
$\sigma=133$~km/s.

\section{Results for the Galaxy}

For model A the expected ratio of Be-X binaries with NSs to the ones
with BHs, $F_{\rm NS/BH}$ was found to be $\sim 7$. For, more
physically realistic, models B and C, this ratio was found to be,
respectively, 27 and 54. This relatively high ratio (for models B
and C) is a natural outcome of {\em (1)} the stellar initial mass
function that provides more neutron stars than black holes and {\em
(2)} common envelope evolution (i.e.~a major mechanism involved in
the formation of interacting binaries) that naturally selects
progenitors of Be X-ray binaries with neutron stars (comparable mass
binaries have more likely survival probabilities) against ones with
black holes (much more likely common envelope mergers).

The expected distributions of orbital periods and eccentricities for
Be/NS and Be/BH binaries for model C are shown in Fig. 1. For
comparison, the observed orbital periods distribution for 27
galactic Be/NS binaries is shown in Fig. 2.

More detailed description of the results is given in Belczy\'nski \&
Zi\'o{\l}kowski (2009).

Now, let me remind that in the Galaxy there are $72$ Be X-ray
binaries known to-date. Out of those, $48$ host a confirmed neutron
star. None, so far, is known to host a black hole. The stellar
population synthesis calculations presented above predict that the
ratio of the binaries with neutron stars to the ones with black
holes should be high $F_{\rm NS/BH} \sim 30-50$. A comparison of
this ratio with the number of confirmed Be -- neutron star X-ray
binaries ($48$) indicates that the expected number of Be -- black
hole X-ray binaries is of the order of only $\sim 0-2$. This is
entirely consistent with the observed Galactic sample. Therefore,
there is no problem of the missing Be+BH X-Ray Binaries for the
Galaxy.

\section{Results for the Magellanic Clouds}

In the Magellanic Clouds there are $98$ Be X-ray binaries known
to-date. Out of those, $63$ host a confirmed neutron star. Again,
none hosts a black hole. The preliminary stellar population
synthesis calculations carried out specifically for the Magellanic
Clouds (Zi\'o{\l}kowski \&Belczy\'nski 2010) predict that the ratio
of the Be X-ray binaries with neutron stars to the ones with black
holes is only $F_{\rm NS/BH} \sim 10$ (for model C). This value is
rather too low, as it implies the expected number of Be+BH X-ray
binaries of the order of $\sim 6$, while none is observed.
Obviously, in contrast to the Galaxy, there is a problem of the
missing Be+BH X-Ray Binaries for the Magellanic Clouds. We believe,
that to remove the discrepancy, one has to take into account a
different history of the star formation rate in the Magellanic
Clouds, with the respect to the Galaxy. During our preliminary
calculations, we used the galactic scenario for the star formation
rate. However, Magellanic Clouds are in many ways very different
from the Galaxy when comparing the population of XRBs. Let us
compare the numbers for three classes of XRBs:

(1) Be XRBs: 72 in the Galaxy vs 98 in the Magellanic Clouds

(2) other High Mass XRBs: 46 vs 11

(3) Low Mass XRBs: 197 vs 2

It is obvious that formation of stars in Magellanic Clouds had to
proceed in a completely different way from that in the Galaxy. We
are currently carrying out the new stellar population synthesis
calculations trying to take into account this fact.


\begin{figure}[h!]
\epsfysize=9.92cm \hspace{1.5cm}\epsfbox{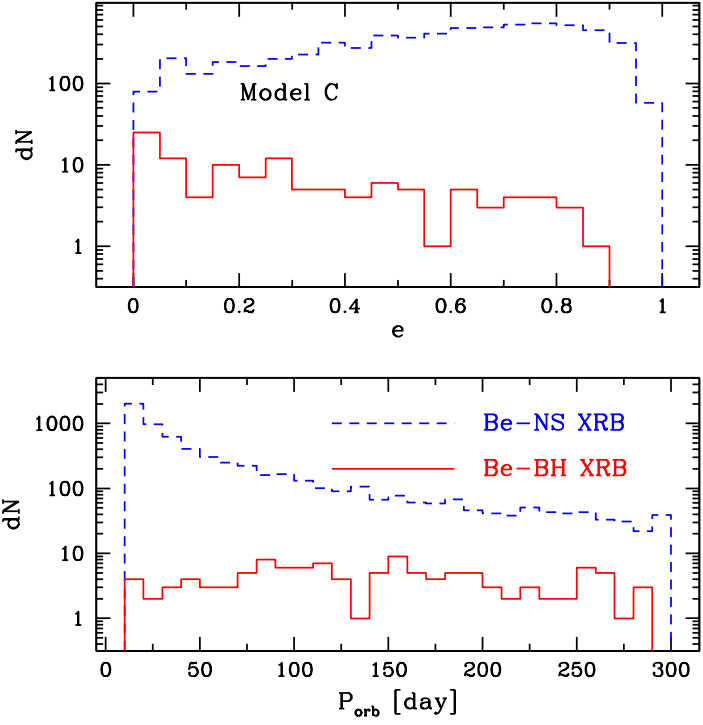}

\caption{Expected orbital period and eccentricity distributions for
Be/NS (broken line) and Be/BH (continuous line) binaries for model C
of stellar population synthesis (see the text).}

\end{figure}


\begin{figure}[h!]

\vspace{-20mm} \epsfysize=6.84cm \hspace{1.6cm}\epsfbox{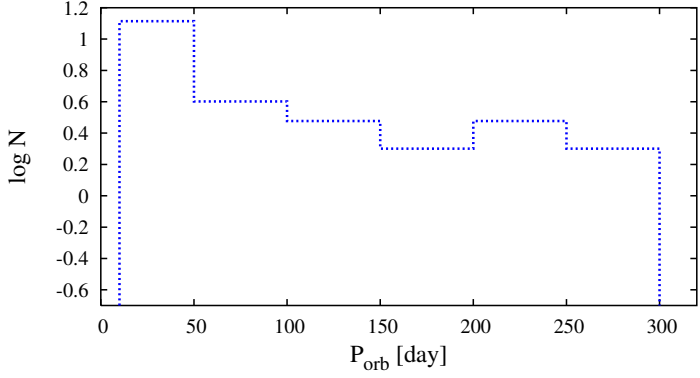}

\caption{Observed orbital period distribution for 27 galactic Be/NS
binaries.}
\end{figure}


\moveright 42mm \hbox{\bf Tab. 1 $-$ Galactic Be X-ray Binaries$^a$}
\nobreak \vspace{5mm}

\moveleft 12mm \vbox{

\begin{tabular}{|c|r|r|c|r|l|r|}
\hline
&&&&&&\\
\multicolumn{1}{|c|}{}&\multicolumn{1}{|r|}{P$_{orb}$}&\multicolumn{1}{|r|}{P$_{spin}$}
&\multicolumn{1}{|c|}{$e$}&
\multicolumn{1}{|r|}{$L_{\rm x,max}^{\rm b}$}&\multicolumn{1}{|l|}{Spectral}&\multicolumn{1}{|r|}{}\\
\multicolumn{1}{|c|}{Name}&\multicolumn{1}{|r|}{[d]}&\multicolumn{1}{|r|}{[s]}&\multicolumn{1}{|c|}{}&\multicolumn{1}{|r|}{[erg/s]}&
\multicolumn{1}{|l|}{type}&\multicolumn{1}{|r|}{Ref$^c$}\\
&&&&&&\\
\hline
&&&&&&\\

2S 0053+604  & 203.59 &                & 0.26 & $3.9\times10^{34}$    & B0.5 Ve &  \\

4U 0115+634  & 24.3 & 3.61               & 0.34 & $3.0\times10^{37}$    & B0.2 Ve &  \\

IGR J01363+6610  & 160 &                &  & $1.3\times10^{35}$    & B1 Ve & 1 \\

RX J0146.9+6121  &  & 1404.2               &  & $3.5\times10^{35}$    & B1 Ve &  \\

IGR J01583+6713  &  &                &  &     & Be &  \\

1E 0236.6+6100  & 26.496 &                & 0.55 & $2.0\times10^{34}$    & B0 Ve &  \\

V 0332+53  & 34.25 & 4.4               & 0.37 & $> 1.0\times10^{38}$    & O8.5 Ve &  \\

4U J0352+309  & 250.3 & 837.0               & 0.11 & $3.0\times10^{35}$    & O9.5 IIIe - B0 Ve &  \\

RX J0440.9+4431  & 155 & 202.5               &  & $3.0\times10^{34}$    & B0.2 Ve & 2 \\

EXO 051910+3737.7  &  &                &  & $1.3\times10^{35}$    & B0 IVpe &  \\

1A J0535+262  & 111.0 & 103.4               & 0.47 & $2.0\times10^{37}$    & O9.7 IIIe &  \\

1H 0556+286  &  &                &  &     & B5ne &  \\

IGR J06074+2205  &  &                &  &     & B0.5 Ve &  \\

HESS J0632+057  & $\sim$ 320 &               &  &     & Be & 3 \\

SAX J0635.2+0533  & 11.2 & 0.0338               &  & $9\div35\times10^{33}$    & B1 IIIe - B2 Ve &  \\

XTE J0658-073  &  & 160.4               &  & $6.6\times10^{36}$    & O9.7 Ve &  \\

3A J0726-260  & 34.5 & 103.2               &  & $2.8\times10^{35}$    & O8-9 Ve &  \\

1H 0739-529  &  &                &  &     & B7 IV-Ve &  \\

1H 0749-600  &  &                &  &     & B8 IIIe &  \\

RX J0812.4-3114  & 81.3 & 31.8851               &  & $1.1\times10^{36}$    & B0.2 IVe &  \\

GS 0834-430  & 105.8 & 12.3               & 0.12 & $1.1\times10^{37}$    & B0-2 III-Ve &  \\

GRO J1008-57  & 247.5 & 93.5               & 0.66 & $2.9\times10^{35}$    & B0e &  \\

RX J1037.5-5647  &  & 862.0               &  & $4.5\times10^{35}$    & B0 III-Ve &  \\

1A 1118-615  &  & 407.68               &  & $5.0\times10^{36}$    & O9.5 Ve &  \\

IGR J11305-6256  &  &                &  & $1\times10^{35}$    & B0.5 IIIe & 4 \\

IGR J11435-6109  & 52.46 & 161.76               &  &     & Be &  \\

2S 1145-619  & 187.5 & 292.4               & $>$ 0.5 & $7.4\times10^{34}$    & B0.2 IIIe &  \\

1H 1253-761  &  &                &  &     & B7 Vne &  \\

1H 1255-567  &  &                &  &     & B5 Ve &  \\

4U 1258-61  & 132.5 & 272.0               & $>$ 0.5 & $1.0\times10^{36}$    & B0.7 Ve &  \\

2RXP J130159.6-635806  &  & 704.0               &  & $5.0\times10^{35}$    & Be ? &  \\

SAX J1324.4-6200  &  & 170.84               &  & $\sim 10^{34}$    & Be ? &  \\

\end{tabular}}

\moveright 30mm \hbox{\bf Tab. 1 $-$ Galactic Be X-ray Binaries$^a$
(continued)} \nobreak \vspace{5mm}

\moveleft 12mm \vbox{
\begin{tabular}{|c|r|r|c|r|l|r|}
\hline
&&&&&&\\
\multicolumn{1}{|c|}{}&\multicolumn{1}{|r|}{P$_{orb}$}&\multicolumn{1}{|r|}{P$_{spin}$}
&\multicolumn{1}{|c|}{$e$}&
\multicolumn{1}{|r|}{$L_{\rm x,max}^{\rm b}$}&\multicolumn{1}{|l|}{Spectral}&\multicolumn{1}{|r|}{}\\
\multicolumn{1}{|c|}{Name}&\multicolumn{1}{|r|}{[d]}&\multicolumn{1}{|r|}{[s]}&\multicolumn{1}{|c|}{}&\multicolumn{1}{|r|}{[erg/s]}&
\multicolumn{1}{|l|}{type}&\multicolumn{1}{|r|}{Ref$^c$}\\
&&&&&&\\
\hline
&&&&&&\\

1WGA J1346.5-6255  &  &                &  & $6.6\times10^{32}$    & B0.5 Ve &  \\

2S 1417-624  & 42.12 & 17.6               & 0.446 & $8.0\times10^{36}$    & B1 Ve &  \\

SAX J1452.8-5949  &  & 437.4               &  & $8.7\times10^{33}$    & Be ? &  \\

XTE J1543-568  & 75.56 & 27.12               & $<$ 0.03 & $> 1.0\times10^{37}$    & Be ? &  \\

2S 1553-542  & 30.6 & 9.26               & $<$ 0.03 & $7.0\times10^{36}$    & Be ? &  \\

IGR J15539-6142  &  &                &  & $3.3\times10^{33}$    & B2-3 Vne &  \\

IGR J16207-5129  &  &                &  & $1.3\times10^{34}$    & B8 IIIe &  \\

SWIFT J1626.6-5156  &  & 15.37               &  &     & B0-2 Ve  & 5 \\

AX J170006-4157  &  & 714.5               &  & $7.2\times10^{34}$    & Be ? &  \\

AX J1700.2-4220  & 44.12  & 54.22    &  &  & Be ? & 6 \\

XTE J1716-379  &  & 670.46    &  &  & Be ? & 7 \\

RX J1739.4-2942  &  &                &  &     & Be ? &  \\

RX J1744.7-2713  &  &                &  & $1.8\times10^{32}$    & B0.5 V-IIIe  &  \\

AX J1749.1-2733   & 185.5  & 66    &  & $3\times10^{36}$ & Be ? & 4,8 \\

AX J1749.2-2725  &  & 220.38               &  & $2.6\times10^{35}$    & Be ? &  \\

GRO J1750-27  & 29.8 & 4.45               &  &     & Be ? &  \\

1XMM J180816.8-191940  &  &                &  & $1.3\times10^{33}$    & Be ? &  \\

AX J1820.5-1434  &  & 152.26               &  & $9.0\times10^{34}$    & O9.5 - B0 Ve &  \\

XTE J1824-141  &  & 120.0               &  &     & Be ? &  \\

1XMM J183327.7-103523  &  &                &  & $1.6\div7.5\times10^{32}$    & B0.5 Ve &  \\

1XMM J183328.7-102409  &  &                &  & $3.3\times10^{32}$    & B1-1.5 IIIe &  \\

GS J1843+00  &  & 29.5               &  & $3.0\times10^{37}$    & B0-2 IV-Ve &  \\

2S 1845-024  & 242.18 & 94.8               & 0.88 & $6.0\times10^{36}$    & Be ? &  \\

IGR J18483-0311  & 18.52  & 21.0526                &  &  $7.8\times10^{36}$   & Be ? & 4,9 \\

XTE J1858+034  &  & 221.0               &  &     & Be ? &  \\

XTE J1859+083  & 60.65  & 9.801               &  &     & Be ? & 10 \\

4U 1901+03  & 22.58 & 2.763               & 0.036 & $1.1\times10^{38}$    & Be ? &  \\

XTE J1906+09  & 28.0 ? & 89.17               &  &     & Be ? &  \\

IGR J19294+1816  & 117.2 & 12.44               &  &     & Be ? & 11 \\

1H 1936+541  &  &                &  &     & Be &  \\

XTE J1946+274  & 169.2 & 15.8               & 0.33 & $5.4\times10^{36}$    & B0-1 IV-Ve &  \\

KS 1947+300  & 40.415 & 18.76               & 0.03 & $2.1\times10^{37}$    & B0 Ve &  \\

\end{tabular}}

\moveright 30mm \hbox{\bf Tab. 1 $-$ Galactic Be X-ray Binaries$^a$
(continued)} \nobreak \vspace{5mm}

\moveleft 12mm \vbox{
\begin{tabular}{|c|r|r|c|r|l|r|}
\hline
&&&&&&\\
\multicolumn{1}{|c|}{}&\multicolumn{1}{|r|}{P$_{orb}$}&\multicolumn{1}{|r|}{P$_{spin}$}
&\multicolumn{1}{|c|}{$e$}&
\multicolumn{1}{|r|}{$L_{\rm x,max}^{\rm b}$}&\multicolumn{1}{|l|}{Spectral}&\multicolumn{1}{|r|}{}\\
\multicolumn{1}{|c|}{Name}&\multicolumn{1}{|r|}{[d]}&\multicolumn{1}{|r|}{[s]}&\multicolumn{1}{|c|}{}&\multicolumn{1}{|r|}{[erg/s]}&
\multicolumn{1}{|l|}{type}&\multicolumn{1}{|r|}{Ref$^c$}\\
&&&&&&\\
\hline
&&&&&&\\

W63 X-1  &  & 36.0               &  &     & Be ? &  \\

EXO 2030+375  & 46.02 & 41.8               & 0.41 & $1.0\times10^{38}$    & B0 Ve &  \\

RX J2030.5+4751  &  &                &  & $1.7\times10^{33}$    & B0.5 V-IIIe  &  \\

GRO J2058+42  & 55.03 & 198.0               &  & $2.0\times10^{36}$    & O9.5-B0 IV-Ve &  \\

SAX J2103.5+4545  & 12.68 & 358.61               & $\sim$ 0.4 & $3.0\times10^{36}$    & B0 Ve &  \\

1H 2138+579  &  & 66.33               &  & $9.1\times10^{35}$    & B0-2 IV-Ve &  \\

1H 2202+501  &  &                &  &     & Be &  \\

SAX J2239.3+6116  & 262.6 & 1247.0               &  & $\sim 2.3\times10^{36}$    & B0 V - B2 IIIe &  \\

&&&&&&\\
\hline
\end{tabular}}

\vspace{5mm} \nopagebreak

{\small $^a$Note that I have not included into the table the
following systems:
 1H 1249-637 (compact companion is probably a white dwarf, Liu et al. 2006); IGR J16318-4848 (B[e] optical component (not Be), Liu et al. 2006);
 1H 1555-552 (Herbig Ae/Be optical component (not Be), Liu et al. 2006); PSR B1259 (non-accreting XRB, Tavani \& Arons 1997)}
\vspace{2mm} \nopagebreak

{\small $^b$Maximum X-ray luminosity}

\vspace{2mm} \nopagebreak

{\small $^c$References for most of the systems are given in
Belczy\'nski \& Zi\'o{\l}kowski 2009; Additional references are: (1)
Corbet \& Krimm 2010; (2) Tsygankov et al. 2011; (3) Falcone et al.
2011 (4) Bird et al. 2010; (5) Nespoli et al. 2011 (6) Markwardt et
al. 2010; (7) Markwardt et al. 2009; (8) Zurita \& Chaty 2008; (9)
Sguera et al. 2007; (10) Corbet et al. 2009b; (11) Corbet \& Krimm
2009.}

\label{galactic}


\vspace{5mm}

\moveright 30mm \hbox{\bf Tab. 2 $-$ Be X-ray Binaries in Magellanic
Clouds$^a$} \nobreak \vspace{5mm}

\moveleft 4mm \vbox{
\begin{tabular}{|c|r|r|r|l|r|}
\hline
&&&&&\\
\multicolumn{1}{|c|}{}&\multicolumn{1}{|r|}{P$_{orb}$}&\multicolumn{1}{|r|}{P$_{spin}$}&
\multicolumn{1}{|r|}{$L_{\rm x,max}^{\rm b}$}&\multicolumn{1}{|l|}{Spectral}&\multicolumn{1}{|r|}{}\\
\multicolumn{1}{|c|}{Name}&\multicolumn{1}{|r|}{[d]}&\multicolumn{1}{|r|}{[s]}&\multicolumn{1}{|r|}{[erg/s]}&
\multicolumn{1}{|l|}{type}&\multicolumn{1}{|r|}{Ref$^c$}\\
&&&&&\\
\hline
&&&&&\\

RX J0032.9-7348  &  &                 & $1.3\times10^{37}$    & Be & 1 \\

RX J0045.6-7313  &  &                 & $1.2\times10^{35}$    & Be ? & 1 \\

RX J0047.3-7312  & 49.06  & 262.23                & $1.8\times10^{36}$    & B0.5e & 1,2,3,4,5 \\

XMMU J004814.1-731003  & 22.5 & 25.55               & $2.1\times10^{35}$    & B1.5e & 1,2,4,5 \\

AX J0048.2-7309  &  &                 & $5.2\times10^{35}$    & Be  & 1,6 \\

RX J0048.5-7302  &  &                 & $3.0\times10^{35}$    & B1.5e  & 1,5,6 \\

AX J0049-729  & 33.37 & 74.676               & $7.5\times10^{36}$    &$\sim$B3 Ve  & 1,3,4 \\
&642 ?&&&&\\

RX J0049.2-7311  & 80.1  &  9.1321            & $3.3\times10^{35}$    & B1-3 IV-Ve  & 1,4,5,7 \\

CXOU J004929.7-731058  &   & 894.36              & $9.3\times10^{34}$    & B1e   & 1,5,8 \\

RX J0049.5-7310  & 91.5  &                & $4.1\times10^{35}$    & Be  & 1,6 \\

RX J0049.5-7331  &  &                 & $5.1\times10^{35}$    & Be  & 1,6 \\

RX J0049.7-7323  & 391 & 746.24               & $7.7\times10^{35}$    & O9.5-B0 III-Ve  & 1,2,3,4,7,8 \\

2S J0050-727  & 44.92  & 7.7912                & $6\times10^{37}$    & B1-1.5 IV-Ve & 1,2,3,4,7 \\

RX J0050.7-7316  & 116.6 & 317.26               & $1.8\times10^{36}$    & B0.5e  & 1,2,3,5,8 \\

RX J0050.7-7332  &  &                & $2.4\times10^{34}$    & Be  & 1,6 \\

RX J0050.9-7310  &  &                & $4.5\times10^{35}$    & B0.5e  & 1,5 \\

2E 0051.1-7304  &  &                & $1.6\times10^{35}$    & B0e  & 1,6 \\

RX J0051.3-7216  & 88.3 & 91.12               & $2.9\times10^{37}$    & B0.5e  & 1,3,4,5,7,9 \\
&115 ?&&&&\\

RX J0051.3-7250  &  &                & $3.6\times10^{34}$    & Be  & 1,6 \\

IGR J00515-7328  &  &                & $1.1\times10^{36}$    & Be  & 10,11 \\

RX J0051.8-7231  & 28.51 & 8.899               & $1.4\times10^{36}$    & O9.5-B0 IV-Ve  & 1,3,4,7,8 \\

RX J0051.8-7310  & 67.88 & 172.4               & $5.6\times10^{36}$    & B0e  & 1,3,4,5,7 \\

RX J0051.9-7255  &  &                & $6.0\times10^{34}$    & Be  & 1,6 \\

XTE J0052-723  & 23.9 ? & 4.78               & $7.2\times10^{37}$    & B0-1 Ve  & 1,7 \\

XTE J0052-725  & 171 & 82.46               & $3.4\times10^{36}$    & B1-3e  & 1,3,4,5,7 \\

RX J0052.1-7319  & 74.51  & 15.3               & $1.3\times10^{37}$    & O9.5-B0 III-Ve  & 1,3,4,7 \\

XMMU J005252.1-721715  & 45.9 & 326.79               & $1\times10^{37}$    & Be & 1,2,4,8,12 \\

RX J0052.9-7158  & 68.54 & 164.7               & $2.0\times10^{37}$    & B0-1 III-Ve  & 1,3,7,13 \\

H 0053-739  & 18.58  & 2.374               & $4.7\times10^{38}$    & O9.5 III-Ve  & 1,3,4,14 \\

CXOU J005323.8-722715  & 143.1 ? & 139.136               & $1.18\times10^{35}$    & B0.5e & 1,2,3,4,5,7 \\

\end{tabular}}

\moveright 18mm \hbox{\bf Tab. 2 $-$ Be X-ray Binaries in Magellanic
Clouds$^a$ (continued)} \nobreak \vspace{5mm}

\moveleft 2mm \vbox{
\begin{tabular}{|c|r|r|r|l|r|}
\hline
&&&&&\\
\multicolumn{1}{|c|}{}&\multicolumn{1}{|r|}{P$_{orb}$}&\multicolumn{1}{|r|}{P$_{spin}$}&
\multicolumn{1}{|r|}{$L_{\rm x,max}^{\rm b}$}&\multicolumn{1}{|l|}{Spectral}&\multicolumn{1}{|r|}{}\\
\multicolumn{1}{|c|}{Name}&\multicolumn{1}{|r|}{[d]}&\multicolumn{1}{|r|}{[s]}&\multicolumn{1}{|r|}{[erg/s]}&
\multicolumn{1}{|l|}{type}&\multicolumn{1}{|r|}{Ref$^c$}\\
&&&&&\\
\hline
&&&&&\\

1WGA J0053.8-7226  & 136.4 & 46.63               & $7.4\times10^{36}$    & O9.5-B1 IV-Ve  & 1,3,4,7 \\

XMMU J005403.8-722632  &  & 341.87               & $1.5\times10^{35}$    & Be ? & 1,2 \\

2E0054.4-7237  & 197  &    140.1            & $4.0\times10^{34}$    & B1 Ve  & 1,3,4 \\

CXOU J005446.3-722523  &  & 4693               & $9\times10^{32}$    & B0-1 Ve  & 8,15 \\

RX J0054.5-7228  &  &                & $1.5\times10^{36}$    & Be  & 1,6 \\

AX J0054.8-7244  & 272   & 497.5               & $5.5\times10^{35}$    & B1 III-Ve  & 1,2,3,4,7 \\

XTE J0055-724  & 62.1 & 58.858               & $4.3\times10^{37}$    & O9 Ve  & 1,2,3,4,7 \\

XTE J0055-727  & 17.95 & 18.3814               & $2.6\times10^{37}$    & B0-2 Ve  & 1,2,4,6,16 \\

XMMU J005517.9-723853  & 412 & 701.7               & $4\times10^{35}$    & O9.5 Ve  & 1,3,4,7 \\

XMMU J005535.2-722906  & 135.3 ? & 644.55               & $4.9\times10^{35}$    & B0-0.5 III-Ve  & 1,2,4 \\

RX J0055.4-7210  & 598 ? &    34.08            & $1.1\times10^{35}$    & B2-3 IV-Ve  & 1,3,4,17 \\

XMMU J005615.2-723754  &  &                & $4.9\times10^{34}$    & Be   & 1,6 \\

CXOU J005736.2-721934  & 152.4 & 564.83               & $1.2\times10^{36}$    & B0-2 IV-Ve & 1,3,4,6,7 \\
&95.3 ? &&&&\\

AX J0057.4-7325  & 21.95 &    101.45            & $1.2\times10^{36}$    & B3-5 Ib-IIe ?  & 1,3,4,7 \\

RX J0057.8-7207  &  &    152.10            & $4.3\times10^{35}$    & B1-2.5 III-Ve   & 1,3,7 \\

RX J0057.8-7202  & 126.4 &    281.1            & $1.6\times10^{36}$    & B0-2 III-Ve   & 1,3,4,7 \\

RX J0057.9-7156  &  &                & $5.7\times10^{34}$    & Be   & 1,6 \\

RX J0058.2-7231  & 59.77 &    291.327            & $2.1\times10^{35}$    & B2-3 Ve   & 1,2,4 \\

RX J0059.2-7138  & 82.37 &    2.7632            & $5.0\times10^{37}$    & B1-1.5 II-IIIe   & 1,3,4,7 \\

XMMU J005929.0-723703  & 224 & 202.52               & $4.5\times10^{36}$    & B0-5 IIIe   & 1,2,4 \\

RX J0059.3-7223  & 71.98 &    200.50            & $3.2\times10^{35}$    & B0-1 Ve   & 1,2,3,4,7 \\

XMMU J010030.2-722035  &  &                & $2.6\times10^{34}$    & Be   & 1,6 \\

RX J0101.0-7206  & 344 ? &    304.49            & $1.3\times10^{36}$    & B0-2 III-Ve   & 1,3,4,7,18 \\

RX J0101.3-7211  & 74.96 &    452.2            & $7.3\times10^{35}$    & B0.5-2 IV-Ve   & 1,3,4,7 \\

RX J0101.6-7204  &  &               & $3.8\times10^{35}$    & Be   & 1,6 \\

AX J0101.8-7223  &  &               & $2.2\times10^{35}$    & Be   & 1,6 \\

CXOU J010206.6-714115  & 101.94 ? & 966.97               & $6.0\times10^{35}$    & B0-0.5 III-Ve & 1,2,4,19 \\
&267.38 ? &&&&\\

RX J0103-722  & 94.4  &   341.87            & $1.5\times10^{36}$
& B0.5 IV-Ve   & 1,2,3,4,7 \\

XTE J0103-728  & 110.0  &   6.8482            & $3.8\times10^{37}$    & O9.5-B0 IV-Ve   & 1,3,4,7 \\
& 24.82 ? &&&&\\

\end{tabular}}

\moveright 18mm \hbox{\bf Tab. 2 $-$ Be X-ray Binaries in Magellanic
Clouds$^a$ (continued)} \nobreak \vspace{5mm}

\moveleft 2mm \vbox{
\begin{tabular}{|c|r|r|r|l|r|}
\hline
&&&&&\\
\multicolumn{1}{|c|}{}&\multicolumn{1}{|r|}{P$_{orb}$}&\multicolumn{1}{|r|}{P$_{spin}$}&
\multicolumn{1}{|r|}{$L_{\rm x,max}^{\rm b}$}&\multicolumn{1}{|l|}{Spectral}&\multicolumn{1}{|r|}{}\\
\multicolumn{1}{|c|}{Name}&\multicolumn{1}{|r|}{[d]}&\multicolumn{1}{|r|}{[s]}&\multicolumn{1}{|r|}{[erg/s]}&
\multicolumn{1}{|l|}{type}&\multicolumn{1}{|r|}{Ref$^c$}\\
&&&&&\\
\hline
&&&&&\\

1H 0103-762  &  &                & $4.3\times10^{35}$    & Be   & 1,6 \\

RX J0103.6-7201  & 26.16 ? &   1323.2            & $6.4\times10^{36}$    & B0 III-Ve   & 1,3,6 \\

RX J0104.1-7243  &  &               & $3.8\times10^{34}$    & Be   & 1,6 \\

RX J0104.5-7221  &  &               & $4.8\times10^{34}$    & Be   & 1,6 \\

AX J0105-722  & 11.09  & 3.343              & $1.5\times10^{35}$    & B1-2 III-Ve   & 1,3,4,6 \\

IGR J01054-7253  & 36.3 & 11.483              &     & Be   & 20,21,22 \\

RX J0105.9-7203  &  & 726              & $1.6\times10^{35}$    & B0.5-3e   & 1,4,23 \\

RX J0106.2-7205  &  &               & $5\times10^{34}$    & B2-5 III-Ve   & 1,6 \\

CXOU J010712.6-723533  & 110.6 & 65.95              & $3.0\times10^{36}$    & B1-1.5 II-IIIe   & 1,2,3 \\

AX J0107.2-7234  &  &               & $2.3\times10^{34}$    & Be   & 6 \\

XTE J00111.2-7317  & 90.5 & 31.0294               & $2.0\times10^{38}$    & O9.5-B1 Ve  & 1,3,4,6 \\

RX J00117.6-7330  &  & 22.07               & $1.2\times10^{38}$    & O9.5-B0 III-Ve  & 1,3 \\

XTE J00119-731  &  & 2.1652               & $6.3\times10^{36}$    & Be ?  & 1,6,7 \\

RX J00119.6-7330  &  &                & $1.5\times10^{34}$    & Be  & 1,6 \\

SMC SXP7.92  &  &  7.92              &     & Be  & 24 \\

XTE SMC 95  & 283 ? & 95.2                & $2\times10^{37}$    & Be ?  & 1,7 \\
&71.3 ? &&&&\\

XTE SMC 144  & 59.38  & 144.1               &     & Be ?  & 1,7 \\

SMC SXP707  & 37.15  &  707.464              & $1.4\times10^{35}$    & Be ?  & 25 \\

SMC SXP723  &   &  723              & $> 3\times10^{37}$    & Be ?  & 26 \\

RX J0209.6-7427  & 40 ? &               & $1.0\times10^{38}$    & B0-1.5 IV-Ve   & 6,27 \\

Swift J045106.8-694803  & 21.64 & 187                &     & Be ?  & 28 \\

IGR J05007-7047  & 30.77 &                & $4.5\times10^{36}$    & B2 IIIe   & 29,30,31 \\

RX J0501.6-7034  &  &                & $7\times10^{34}$    & B0 Ve   & 1,6 \\

RX J0502.9-6626  &  & 4.0635               & $4\times10^{37}$    & B0 Ve   & 1,6 \\

Swift J0513.4-6547  &  & 27.28                &     & Be ?  & 32 \\

RX J0516.0-6916  &  &                & $5\times10^{35}$    & B1 Ve    & 1,6,33 \\

XMMU J052016.0-692505  &  &                & $1\times10^{38}$    & B0-3e   & 1,34 \\

RX J0520.5-6932  & 24.4 &                & $8\times10^{38}$    & O9 Ve    & 1 \\

RX J0529.8-6556  &  & 69.5               & $2\times10^{36}$    & B0.5 Ve    & 1 \\

RX J0530.1-6551  &  & 271.97               & $3.9\times10^{35}$    & Be ?   & 1,35,36 \\

EXO 053109-6609.2  & 25.4 & 13.7               & $1\times10^{37}$    & B0.7 Ve    & 1,6 \\

\end{tabular}}

\moveright 18mm \hbox{\bf Tab. 2 $-$ Be X-ray Binaries in Magellanic
Clouds$^a$ (continued)} \nobreak \vspace{5mm}

\moveleft 7mm \vbox{
\begin{tabular}{|c|r|r|r|l|r|}
\hline
&&&&&\\
\multicolumn{1}{|c|}{}&\multicolumn{1}{|r|}{P$_{orb}$}&\multicolumn{1}{|r|}{P$_{spin}$}&
\multicolumn{1}{|r|}{$L_{\rm x,max}^{\rm b}$}&\multicolumn{1}{|l|}{Spectral}&\multicolumn{1}{|r|}{}\\
\multicolumn{1}{|c|}{Name}&\multicolumn{1}{|r|}{[d]}&\multicolumn{1}{|r|}{[s]}&\multicolumn{1}{|r|}{[erg/s]}&
\multicolumn{1}{|l|}{type}&\multicolumn{1}{|r|}{Ref$^c$}\\
&&&&&\\
\hline
&&&&&\\

RX J0531.5-6518  &  &                & $3\times10^{35}$    & B2 Ve   & 1,37,38 \\

RX J0532.4-6535  &  &                & $7\times10^{34}$    & Be ?   & 1,37 \\

RX J0535.0-6700  & 241 ? &                & $3\times10^{35}$    & B0 Ve    & 1,37,38 \\

1A 0535-668  & 16.70 & 0.069               & $1\times10^{39}$    & B0.5 IIIe   & 1,6,39 \\
&16.651 &&&&\\

XMMU J054134.7-682550  &   & 61.601               & $2.0\times10^{38}$    & Be ?   & 1,40 \\

H 0544-665  &  &                & $1\times10^{37}$    & B0 Ve   & 1,6 \\

1SAX J0544.1-7100  & 286 ? & 96.08               & $2\times10^{36}$    & B0 Ve   & 1,6 \\

IGR J05414-6858  &  &               & $8\times10^{36}$    & B1-2 IIIe   & 41,42 \\

&&&&&\\
\hline
\end{tabular}}

\vspace{5mm} \nopagebreak

{\small $^a$Note that I included into the table the system XMMU
J052016.0-692505, although it is possible that it contains a white
dwarf component instead of a NS (Kahabka et al. 2006; Raguzova
2008).

I did not include the system MAXI J1836-194, for which the initial
optical spectroscopy suggested a Be optical component  (Cenko et al.
2011), but this was not confirmed by later observations (Rau et al.
2011).

Excentricities are not given for Magellanic Clouds systems, as they
are estimated for only two systems: IGR J01054-7253 ($e = 0.28$) and
1A 0535-668 ($e \ge 0.5$).

 \vspace{2mm} \nopagebreak

{\small $^b$Maximum X-ray luminosity}

\vspace{2mm} \nopagebreak

{\small $^c$ (1) Raguzova 2007; (2) Haberl et al. 2008; (3) McBride
et al. 2008; (4) Rajoelimanana et al. 2011a; (5) Antoniou et al.
2009; (6) Liu et al. 2006; (7) Galache et al. 2008; (8) Laycock et
al. 2010; (9) Coe et al. 2005; (10) Sturm et al. (2011); (11) Kennea
et al. 2011; (12) Coe et al. 2008; (13) Corbet et al. 2004; (14)
Schmidtke et al. 2009a; (15) Schmidtke \& Cowley 2010; (16) Coe \&
Udalski 2008; (17) Haberl \& Eger 2008; (18) Sasaki et al. 2003;
(19) Schmidtke et al. 2009b; (20) Coe et al. 2009; (21) Corbet et
al. 2009a; (22) Townsend et al. 2009; (23) Eger \& Haberl 2008; (24)
Corbet et al. 2008; (25) Rajoelimanana et al. 2011b; (26) Townsend
et al. 2011; (27) Kahabka \& Hilker 2005; (28) Beardmore et al.
2009; (29) Masetti et al. 2006; (30) Bird et al. 2010; (31) La
Parola et al. 2010; (32) Krimm et al. 2009; (33) Schmidtke et al.
1999; (34) Kahabka et al. 2006; (35) Haberl et al. 2003; (36)
Shtykovskiy \& Gilfanov 2005; (37) Haberl \& Pietsch 1999; (38)
Negueruela \& Coe 2002; (39) Alcock et al. 2001; (40) Markwardt et
al. 2007; (41) Grebenev \& Lutovinov 2010; (42) Rau et al. 2010.

\label{MCs}


\vspace{3mm}

\section{Acknowledgements}

This work was partially supported by the Polish Ministry of Science
and Higher Education (MSHE) project 362/1/N-INTEGRAL (2009-2012).

\end{document}